\documentclass[a4paper]{article}

\usepackage{icrc2013}
\usepackage[english]{babel}

\title{The Statistical and Numerical Study of the Longitudinally Asymmetric Distribution of Solar Proton Events Affecting the Earth Environment of 1996-2011}

\shorttitle{Longitudinally Asymmetric Distribution of Solar Proton
Events}

\authors{Hongqing He$^{1,2}$, Weixing Wan$^{1}$}

\afiliations{ $^1$ Beijing National Observatory of Space
Environment, Institute of Geology and Geophysics, Chinese Academy of
Sciences, Beijing 100029, China \\
$^2$ CAS Key Laboratory of Geospace Environment, Department of
Geophysics and Planetary Sciences, University of Science and
Technology of China, Hefei, Anhui 230026, China}

\email{hqhe@mail.iggcas.ac.cn}

\abstract{Large solar proton events (SPEs) affect the
solar-terrestrial space environment and become a very important
aspect in space weather research. In this work, we statistically
investigate 78 solar proton events of 1996-2011 and find that there
exists a longitudinally asymmetric distribution of flare sources of
the solar proton events observed near 1 AU, namely, with the same
longitude separation between magnetic field line footpoint of
observer and flare sources, the number of the solar proton events
originating from sources located at eastern side of the nominal
magnetic footpoint of observer is much larger than that of the solar
proton events originating from sources located at western side. A
complete model calculation of solar energetic particle (SEP)
propagation in the three-dimensional Parker interplanetary magnetic
field is presented to give a numerical explanation for this
longitudinally asymmetric distribution phenomenon. We find that the
longitudinally asymmetric distribution of solar proton events
results from the east-west azimuthal asymmetry in the topology of
the Parker interplanetary magnetic field as well as the effects of
perpendicular diffusion on the transport of SEPs in the heliosphere.
Our results would be valuable in understanding the solar-terrestrial
relations and useful in space weather forecasting.}


\begin{document}
\maketitle

\section{Introduction}
Solar energetic particles (SEPs), which are charged energetic
particles occasionally emitted by the Sun, risk the health of
astronauts working in space and damage electronic components on
satellites, so they have become a very important aspect in affecting
solar-terrestrial space environment and space weather.
Theoretically, SEPs observed in the interplanetary magnetic field
provide fundamental information regarding acceleration mechanisms
and transport processes of charged energetic particles. Therefore,
the subject has become a focus of astrophysics, space physics, and
plasma physics.

Through several decades of investigations in the community with
spacecraft observations and theoretical modeling, significant
progresses have been achieved toward a better understanding of the
transport processes of SEPs. As a pioneering work, \cite{Parker1965}
provided a focused transport equation to investigate the modulation
of galactic cosmic rays and the transport of SEPs. It is very
difficult to analytically solve the multidimensional transport
equation; consequently, numerical calculations are usually adopted
(e.g., \cite{Zhang1999,He2011}).

Recently, \cite{He2011} investigated the effects of particle source
characteristics on SEP observations at 1 AU and found that the
perpendicular diffusion has a very important influence on the
propagation of SEPs in the heliosphere, particularly when a
spacecraft is not directly connected to the acceleration regions
either on the Sun or near the CME-driven shocks by the
interplanetary magnetic field lines; in such cases, the earliest
arriving particles can be seen propagating toward the Sun, having
scattered backward at large distances. \cite{He2012a,He2012b}
presented a direct method to quickly and explicitly determine the
spatially dependent parallel, radial, and perpendicular mean free
paths of SEPs with adiabatic focusing. They reported that when the
turbulence strength $\delta B/B$ is sufficiently large, e.g.,
$\delta B/B\geq2.34$, the ratio
$\lambda_{\perp}/\lambda_{\parallel}$ would approach or exceed
unity.

In this work, we statistically investigate 78 solar proton events
(SPEs) of 1996-2011 observed near 1 AU and find that with the same
longitude separation between magnetic field line footpoint of
observer and flare sources, the number of the solar proton events
originating from sources located at eastern side of the nominal
magnetic footpoint of observer is much larger than that of the solar
proton events originating from sources located at western side. Our
systematical simulations also confirm the longitudinally asymmetric
distribution of the solar proton events observed in the
interplanetary space including Earth's orbit. We find that the
longitudinally asymmetric distribution of solar proton events
results from the east-west azimuthal asymmetry in the geometry of
the Parker interplanetary magnetic field as well as the effects of
perpendicular diffusion on the transport processes of SEPs in the
heliosphere. Our results would be valuable in understanding the
solar-terrestrial relations and useful in space weather forecasting.

\section{Data Set}
We collect 78 solar proton events affecting the Earth environment
observed near 1 AU in the time period 1996-2011 from Space Weather
Prediction Center of NOAA at
http://www.swpc.noaa.gov/ftpdir/indices/SPE.txt. The data set covers
an uninterrupted time period of 16 years including the whole solar
cycle 23 and the rising phase of solar cycle 24. The maximum proton
flux of each particle event measured by GOES spacecraft at
Geosynchronous orbit for energies $>10$ MeV is greater than or equal
to 10 particle flux units (pfu, 1 pfu=1 particles/($cm^{2}-sr-s$)).
Note that the 78 solar proton events are selected from the original
listing by excluding the events without definite location on the
solar surface, since the explicit data of the active regions are
indispensable to the investigations. The solar wind speed used in
this work is averaged from SOHO or ACE data measured within the time
range $[t_{max}-2 hours, t_{max}+2 hours]$, where $t_{max}$ is the
maximum time of each SEP event.

\section{Statistical Results}
The heliospheric magnetic field is ``frozen-into" the solar wind and
travels along with it to form the Archimedes spirals. The well-known
Parker spiral can be expressed as
\begin{equation}
\phi_{s}=\Omega r/V^{sw},\label{spiral-angle}
\end{equation}
where $\Omega$ is the angular rotation rate of the Sun, $r$ is the
heliocentric distance, and $V^{sw}$ is the solar wind speed. For
each solar proton event, by inputting the solar wind speed $V^{sw}$
into Equation (\ref{spiral-angle}), we can derive the Parker spiral
angle $\phi_{s}$ of the nominal magnetic field line connecting the
observer at 1 AU with the Sun. That is to say, we can determine the
heliographic longitude of the nominal magnetic footpoint of the
spacecraft located at Earth's orbit during the SEP events. For
convenience in the investigations, we mark the west and east
heliographic longitudes $\phi$ of the flare sources on the solar
disk as positive and negative values, respectively. Then for each
solar proton event, we can obtain the relatively longitudinal
distance $\phi_{r}$ between the heliographic longitudes of the flare
location and the nominal magnetic footpoint of the spacecraft
located at 1 AU during the particle event, through the relation
\begin{equation}
\phi_{r}=\phi-\phi_{s}=\phi-\Omega r/V^{sw}.
\label{relative-longitude}
\end{equation}

\begin{figure}[t]
  \centering
  \includegraphics[width=0.4\textwidth]{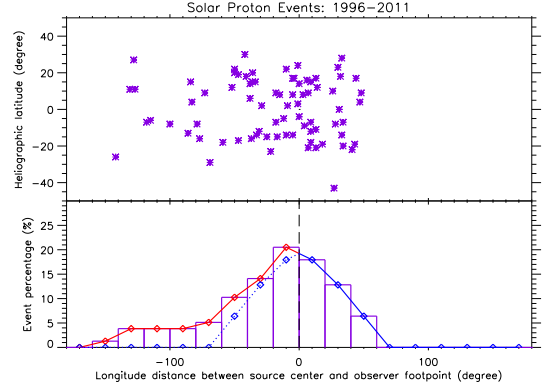}
  \caption{The spatial distribution on the solar surface of the flare
 sources of the 78 solar proton events observed near 1 AU in the time period 1996-2011.
 Upper panel: the latitudinal and longitudinal distribution of the solar sources
 of the proton events. Lower panel: the number percentage distribution
 of the solar events along the relatively longitudinal distance $\phi_{r}$,
 which is divided into 18 bins. With the same longitude separation between
 the magnetic footpoint of observer and solar sources, the number of the solar
proton events originating from sources located at eastern (left)
side of the magnetic footpoint of observer is much larger than that
of the solar proton events originating from sources located at
western (right) side.}
  \label{observation}
\end{figure}

Figure \ref{observation} presents the spatial distribution on the
solar surface of the flare sources of the 78 solar proton events
observed near 1 AU in the time period 1996-2011. Similar to the
heliographic longitude, the heliographic latitude is marked as a
positive value when the flare source located at the northern side of
the solar equator and a negative value when the flare source located
at the southern side. From the upper panel of Figure
\ref{observation}, we can see that the majority of flare sources of
the solar proton events are distributed in the heliographic latitude
range $[S30^{\circ}, N30^{\circ}]$. The lower panel of Figure
\ref{observation} shows the number percentage distribution of the
solar events along the relatively longitudinal distance $\phi_{r}$,
which is divided into 18 bins. The event number percentage $P_{i}$
$(i=1, 2, \ldots, 18)$ within each bin of $\phi_{r}$ is obtained as
follows:
\begin{equation}
P_{i}=\frac{N_{i}}{N}=\frac{N_{i}}{78}\times100\%,
\label{event-percentage}
\end{equation}
where $N_{i}$ $(i=1, 2, \ldots, 18)$ denotes the corresponding solar
event number in each bin. The vertical dashed line in the middle of
the lower panel in Figure \ref{observation} indicates the
$0^{\circ}$ relatively longitudinal distance, where the nominal
magnetic footpoint of the observer and the flare source have the
same heliographic longitude. At the left side of this line, the
negative values indicate that the flare sources locate at the
eastern side of the nominal magnetic footpoint of observer, whereas
the positive values at the right side of the middle line indicate
that the flare sources locate at the western side of the magnetic
footpoint of observer. It can be seen that the event number
percentage is maximum in the interval $[-20^{\circ}, 0^{\circ}]$
among all the 18 relatively longitudinal bins. Interestingly, we
find that the event number percentage distribution is asymmetric
about the vertical dashed line in the middle. We connect the event
number percentages at the central value within each relatively
longitudinal interval with red and blue lines on the left-hand and
right-hand halves, respectively. We further mirror the event number
percentages on the right-hand half to the left-hand half with a blue
dotted line. Then we can clearly see that the event number
percentage within each interval on the left-hand half is
systematically larger than that in the symmetrical interval on the
right-hand half. That is to say, with the same longitude separation
between magnetic field line footpoint of observer and flare sources,
the number of the solar proton events originating from sources
located at eastern side of the nominal magnetic footpoint of
observer is much larger than that of the solar proton events
originating from sources located at western side.

\section{Numerical Model and Results}

\subsection{Model}
In our model, the five-dimensional focussed transport equation that
governs the gyrophase-averaged distribution function
$f(\textbf{x},\mu,p,t)$ of SEPs can be written as (e.g.,
\cite{He2011})
\begin{eqnarray}
{}&&\frac{\partial f}{\partial t}+\mu v\frac{\partial f}{\partial
z}+{\bf V}^{sw}\cdot\nabla f+\frac{dp}{dt}\frac{\partial f}{\partial
p}+\frac{d\mu}{dt}\frac{\partial f}{\partial
\mu}-\frac{\partial}{\partial\mu}\left(D_{\mu\mu}\frac{\partial
f}{\partial \mu}\right)  \nonumber\\
{}&&-\frac{\partial}{\partial x}\bigg(\kappa_{xx}\frac{\partial
f}{\partial x}\bigg)-\frac{\partial}{\partial
y}\left(\kappa_{yy}\frac{\partial f}{\partial y}\right)=Q({\bf
x},p,t),\label{transport-equation}
\end{eqnarray}
where $\textbf{x}$ is particle's position, $z$ is the coordinate
along the magnetic field spiral, $p$ is particle's momentum, $\mu$
is pitch-angle cosine, and $Q$ is source term.

Under the diffusion approximation for a nearly isotropic pitch-angle
distribution, the parallel mean free path $\lambda_{\parallel}$
could be written as (\cite{Jokipii1966})
\begin{equation}
\lambda_{\parallel}=\frac{3v}{8}\int_{-1}^{+1}\frac{(1-\mu^{2})^{2}}{D_{\mu\mu}}d\mu.
\label{parallel-path}
\end{equation}
In this work, we typically use radial mean free path
$\lambda_{r}=0.3$ AU (corresponding to parallel mean free path
$\lambda_{\parallel}=0.67$ AU at 1 AU) and perpendicular mean free
paths $\lambda_{x}=\lambda_{y}=0.006$ AU for 50 MeV protons. We
utilize the time-backward Markov stochastic process method
(\cite{Zhang1999}) to numerically solve the focused transport
equation (\ref{transport-equation}). In our simulations, the Parker
interplanetary magnetic field $\textbf{B}$ is set so that its
magnitude at $1 AU$ is $5 nT$, and the solar wind speed is typically
set as $V^{sw}=400~km~s^{-1}$. Typical sizes of SEP sources (flares
or CMEs) are tens of degrees wide in heliographic latitude and
longitude, so in the simulations, we set SEP sources with limited
coverage of $30^{\circ}$ in latitude and longitude. Generally, the
unit of omnidirectional flux is used as
$cm^{-2}-sr^{-1}-s^{-1}-MeV^{-1}$; in this work, however, we use an
arbitrary unit for convenience in plotting figures.

\subsection{Numerical Results}
We simulate 18 cases to show the effect of solar source location on
the SEP flux observed in the interplanetary space. All the solar
particle sources have the same coverage, but their centers are
located at various heliographic longitudes in the solar equator.
Specifically, the longitude separations between the centers of 18
solar sources and the magnetic footpoint of the observer are set to
be the following values: $0^{\circ}$, $20^{\circ}$ west,
$20^{\circ}$ east, $\ldots$, $160^{\circ}$ west, $160^{\circ}$ east,
$180^{\circ}$ east. All the other conditions and parameters are the
same for each case. In these simulations, we typically investigate
$50$ MeV protons detected at 1 AU, $90^{\circ}$ colatitude.

\begin{figure}[t]
  \centering
  \includegraphics[width=0.4\textwidth]{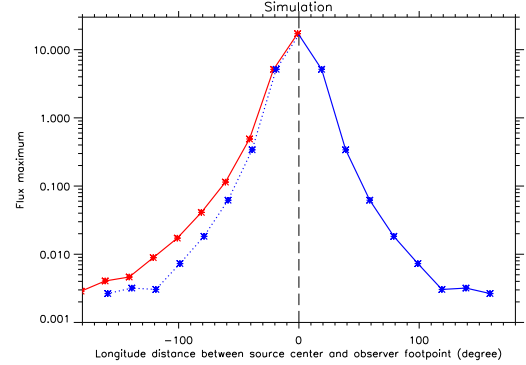}
  \caption{The flux maxima of the SEP events originating from 18 solar sources
with different longitudinal locations relative to the magnetic
footpoint of the observer at 1 AU, $90^{\circ}$ colatitude. With the
same longitude separations between the solar source centers and the
magnetic footpoint of the observer, the flux maxima of the SEP
events originating from solar sources located at eastern (left) side
of the observer footpoint are systematically larger than those of
the SEP events originating from sources located at western (right)
side.}
  \label{simulation}
\end{figure}

Figure \ref{simulation} shows the flux maxima of the SEP events
originating from the 18 solar sources with different longitudinal
locations relative to the magnetic footpoint of the observer at 1
AU, $90^{\circ}$ colatitude. The X-axis in Figure \ref{simulation}
is the relatively longitudinal distance $\phi_{r}$ between the solar
source center and the magnetic footpoint of the observer, same as
that in Figure \ref{observation}. The vertical dashed line in the
middle of Figure \ref{simulation} denotes the $0^{\circ}$ relatively
longitudinal distance. As we can see, the maximum flux of SEPs in
this case is the largest among all the 18 cases investigated. This
means that when the observer in the interplanetary space is
connected directly to the solar source by heliospheric magnetic
field lines, the SEP flux is larger than that otherwise. We connect
the flux maxima of SEPs in the simulated cases on the left-hand and
right-hand halves with red and blue curves, respectively. It can be
obviously seen from Figure \ref{simulation} that, the farther the
solar source center is away from the magnetic footpoint of the
observer, the smaller is the maximum flux of SEPs observed (see also
\cite{He2011}). Similar to Figure \ref{observation}, we further
mirror the flux maxima of SEPs on the right-hand half to the
left-hand half with a blue dotted curve. We can obviously observe
that with the same longitude separations between the solar source
centers and the magnetic footpoint of the observer, the flux maxima
of SEP events originating from solar sources located at eastern side
of the observer footpoint are systematically larger than those of
the SEP events originating from sources located at western side.

\section{Relationship between Simulations and Observations}
Generally, a SEP event with a higher peak flux will reveal a larger
intensity relative to other events through the entire evolution
process including the rising phase and the decay phase (see the
results, such as Figure 2 and Figure 3, in \cite{He2011}).
Generally, the probability of the SEP events being observed by
spacecraft near Earth's orbit depends on the intensities of the
particles in the events. An intense SEP event with high flux will
experience relatively weak influences by the interplanetary
structures (such as magnetic cloud, heliospheric current sheet,
corotating interaction region, etc.) during the propagation process
in the heliosphere. However, the not so intense SEP events with
small fluxes will be significantly affected by the interplanetary
structures and then to weaken or even dissipate before entering
Earth's orbit. Consequently, the intense SEP events would have
larger probability of reaching the Earth and being observed by
spacecraft at 1 AU; whereas the relatively weak SEP events would
have lower probability to arrive at the Earth and to be recorded by
spacecraft.

As shown in Figure \ref{simulation}, the SEP events originating from
solar sources located at eastern side of the magnetic footpoint of
observer reveal larger flux maxima (corresponding to entire fluxes)
than those originating from solar sources located at western side.
Therefore, the SEP events originating from solar sources located at
eastern relative longitudes would have higher probability to reach
the Earth and to be observed by spacecraft near 1 AU. On the
contrary, the SEP events from solar sources on the west side
relative to the magnetic footpoint of observer will be more easily
influenced by the interplanetary structures and then to weaken or
even dissipate during their propagation processes. As a result, the
spacecraft at 1 AU would have a larger probability to miss the SEP
events originating from solar sources located at western relative
longitudes. Accordingly, taking into account a large data set with
78 solar proton events accumulated over a long time period, we can
expect that with the same relatively longitudinal distance, the
number of the solar proton events originating from solar sources
located at eastern side of the magnetic footpoint of observer would
be larger than that of the solar proton events originating from
sources located at western side, just as shown in Figure
\ref{observation}. Therefore, our numerical simulations explain and
confirm the observational results; and the combination of
simulations and spacecraft observations demonstrates that there
exists a longitudinally asymmetric distribution of solar sources of
the proton events observed near 1 AU.

\section{Summary and Discussion}
In this work, we statistically investigate a large data set with 78
solar proton events accumulated over a long and uninterrupted time
period 1996-2011. We find that with the same longitude separation
between magnetic footpoint of observer and solar sources, the number
of the proton events originating from the solar sources located at
eastern side of the magnetic footpoint of observer is systematically
larger than that of the proton events associated with the sources
located at western side. Our simulation results show that with the
same longitude separation between the solar sources and the magnetic
footpoint of the observer, the flux of SEPs released from the solar
source located east is systematically higher than that of SEPs
originating from the source located west. Consequently, the SEP
events originating from solar sources located at the eastern
relative longitudes would be detected and recorded with somewhat
higher probabilities by the observer near 1 AU. Therefore, our
numerical simulations testify and confirm the statistical results of
the solar proton events accumulated over 16 years.

\begin{figure}[t]
  \centering
  \includegraphics[width=0.4\textwidth]{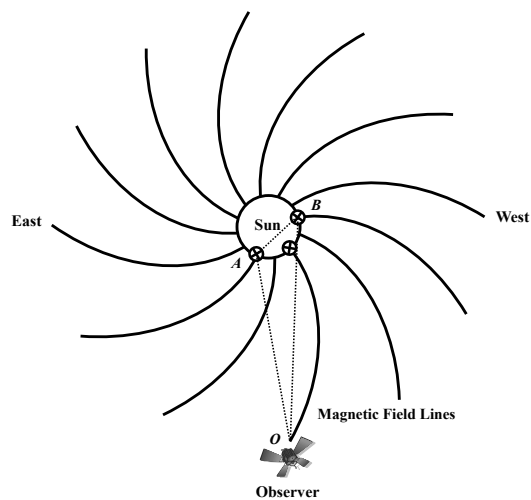}
  \caption{Sketch to show the east-west azimuthal asymmetry in the topology of
the Parker interplanetary magnetic field. With the same longitudinal
separation between the magnetic footpoint of the observer and the
solar sources, the virtual distance $\overline{OA}$ between the
observer (O) in the interplanetary space and the solar source (A) on
the left is shorter than that $\overline{OB}$ between the observer
and the solar source (B) on the right. On account of
$\overline{OA}<\overline{OB}$, the SEPs originating from the solar
sources on the east (left) side relative to the magnetic footpoint
of the observer would have a lower probability to weaken or
dissipate before arriving at the observer.}
  \label{Parker-spiral}
\end{figure}

Generally, as viewed from above the north pole of the Sun, the
Parker spiral magnetic field lines originating from the solar
surface would meander clockwise to somewhere with more eastern
longitudes in the interplanetary space. The scenario can be seen in
Figure \ref{Parker-spiral}. As we can see, there is an east-west
azimuthal asymmetry in the topology of the Parker interplanetary
magnetic field. As a result, with the same longitude separation
between the magnetic footpoint of the observer and the solar
sources, the virtual distance $\overline{OA}$ between the observer
(O) in the interplanetary space and the solar source (A) on the left
is shorter than that $\overline{OB}$ between the observer and the
solar source (B) on the right. According to the theoretical and
observational investigations (e.g., \cite{He2012b}), for physical
conditions representative of the solar wind, the perpendicular
diffusion coefficient would be a few percent of the parallel
diffusion coefficient in magnitude. In general, SEPs would mainly
transport along the average interplanetary magnetic field after they
have been produced from the surface of the Sun or a CME-driven
shock. Therefore, when the observer is connected directly to the
solar source by the interplanetary magnetic field lines, the
omnidirectional flux and peak flux of SEPs are larger than those in
the cases where the observer is not connected directly to the solar
sources. The farther the solar source center is away from the
magnetic footpoint of the observer, the smaller are the
omnidirectional flux and peak flux of SEPs observed. In addition to
the parallel diffusion, the SEPs will experience perpendicular
diffusion to cross the heliospheric magnetic field lines during
their transport processes in the interplanetary space. This is why
the SEP events can be observed by a spacecraft which is not directly
connected to the acceleration regions. On account of the virtual
distance relationship $\overline{OA}<\overline{OB}$, the SEPs
originating from the solar sources located at eastern side of the
magnetic footpoint of observer would have a lower probability to
weaken or dissipate before arriving at the observer. Consequently,
with the same longitudinal distances between the solar sources and
the observer footpoint, the fluxes of the SEP events associated with
solar sources located east are systematically larger than those of
the SEP events associated with sources located west. Therefore, the
longitudinally asymmetric distribution of solar proton events
results from the east-west azimuthal asymmetry in the topology of
the Parker interplanetary magnetic field as well as the effects of
perpendicular diffusion on the transport processes of SEPs in the
heliosphere.

\vspace*{0.5cm} \footnotesize{{\bf Acknowledgment: }{We were
supported partly by grants NNSFC 41204130, NNSFC 41131066, the
National Important Basic Research Project 2011CB811405, the Chinese
Academy of Sciences KZZD-EW-01-2, the China Postdoctoral Science
Foundation, the Open Research Program from Key Laboratory of
Geospace Environment, CAS, and the Key Laboratory of Random Complex
Structures and Data Science, CAS.}}

\end{document}